\documentclass[11pt]{article}

\usepackage[latin1]{inputenc}
\usepackage{amsmath,amssymb,amsthm}
\usepackage{graphicx,graphics,color,epsfig}
\usepackage{hyperref}
\usepackage{dsfont}
\usepackage{cite}

\newcommand{\un}{\textbf{1}\normalsize}

\newcommand{\CA}{\mathcal A}

\newcommand{\CI}{\mathcal I}

\newcommand{\CP}{\mathcal P}

\newcommand{\CT}{\mathcal T}

\newcommand{\CX}{\mathcal X}

\newcommand{\BN}{\mathbb N}

\newcommand{\BR}{\mathbb R}

\theoremstyle{plain}
\newtheorem{theo}{Theorem}
\newtheorem{prop}{Proposition}
\newtheorem{coro}{Corollary}

\theoremstyle{definition}
\newtheorem{defi}{Definition}

\theoremstyle{remark}

\newtheorem{exam}{Example}

\title{\textsc{Performance bounds in wormhole routing}\\
  A Network Calculus Approach}

\author{Nadir Farhi\footnote{supported by ANR Pegase Project}\; and Bruno Gaujal\\ INRIA Rh\^one-Alpes and LIG laboratory,
  Grenoble, France \\~ \vspace{3mm} \small{\texttt{nadir.farhi@inria.fr}}}

\date{}

\begin{document}
\maketitle

\begin{abstract}
We present a model of performance bound calculus on feedforward
networks where data packets are routed under wormhole routing
discipline. We are interested in determining maximum end-to-end
delays and backlogs of messages or packets going from a source
node to a destination node, through a given virtual path in the
network. Our objective here is to give a network calculus approach
for calculating the performance bounds. First we propose a new
concept of curves that we call \emph{packet curves}. The curves
permit to model constraints on packet lengths of a given data
flow, when the lengths are allowed to be different. Second, we use
this new concept to propose an approach for calculating residual
services for data flows served under non preemptive service
disciplines. Third, we model a binary switch (with two input ports
and two output ports), where data is served under wormhole
discipline. We present our approach for computing the residual
services and deduce the worst case bounds for flows passing
through a wormhole binary switch. Finally, we illustrate this approach in
numerical examples, and show how to extend it to feedforward
networks.
\end{abstract}

\textbf{Keywords} Network Calculus, Quality of Service Guarantees,
Wormhole Routing, Spacewire.

\section*{Introduction}

In this article, we present  an approach for end-to-end delay
computation in communication networks, where data messages are routed
under wormhole routing discipline. Wormhole
routing~\cite{LN91,BC93,MXEN93,NM93,SJ96,DT04} is a popular
routing strategy based on known fixed links. It is a subset of
flow control methods called flit-buffer flow
control~\cite{CPC08,Dan94,BBPTG07,Tia97}. Each message is
transmitted as a contiguous sequence of flow control units. The
sequence move from a source node to a destination node in a
pipeline manner, like a burrowing worm. We will be based here on a
spacewire~\footnote{Spacewire~\cite{ESA,NOR08} is a spacecraft
communication network coordinated by the European Space Agency, in
collaboration with other international space agencies including
NASA, JAXA, and RKA. Spacewire is based in part on the IEEE~1355
standard of communication~\cite{IEEE1355}.} implementation of
wormhole routing.

Our approach is based on \emph{network calculus
theory}~\cite{Cha00,BT04,Kum04}. Remarkably, this theory is almost
solely based on two objects, namely {\it arrival curves} and {\it
  service curves} that are used to express constraints on arrival
flows and service capacities.
Performance bounds
are then derived by cleverly handling arrival and service curves, and by
taking into account the service policies. In this article, we consider
the general case where packets of one flow may have different
lengths. We propose a new object, namely  \emph{packet curves}, where
information about packet lengths are summarized in curves, in the
same way as arrival times of data are summarized in  arrival
curves, in classical network calculus theory. In general, when
packets may have different lengths, only the minimum and the
maximum sizes  are taken into account in end-to-end delay
calculus. We show here that the whole available information about
packet lengths, summarized in {packet curves},
can be taken into account in end-to-end delay calculus. In
particular, we show how to compute the \emph{minimum mean of service} and
give the gain of our approach with respect to the existing calculus
approach.

The approach of packet curves is then applied to calculate
residual services of arrival flows routed under wormhole routing
discipline, where packets of each flow may have different lengths,
and where information about the sequence of packet lengths of a
given flow is given by packet curves. We study in detail the
routing on a \emph{wormhole switch}, based  on a
spacewire implementation of wormhole routing, where flows of
different output port destinations may arrive onto one buffered
input port, and where messages served on a given output port may
arrive from several input ports, and are served with round-robin
service policy. We show, on a numerical example, the maximum delay
calculus for a message passing through a spacewire-like switch.
Finally and briefly, we explain the extension of this approach to
feedforward communication networks.

\section{Wormhole routing}

Wormhole routing~\cite{LN91,BC93,MXEN93,NM93,SJ96,DT04} is a routing
strategy used in parallel computers
and with a variety of machines such as Intel data, MIT J machine
and MIT April~\cite{RSSW94}. Unlike in store-and-forward routing, where packets are received, stored, and then routed, in wormhole
routing, packets are routed as follows: Each packet of data contains a header
giving the destination address of the packet. As soon as the
header of a packet is received at an input port of a given switch, the latter determines
the corresponding output port by checking the destination address.
\begin{itemize}
  \item If the requested output port is free, then the packet is routed immediately to that
    output port. Once the packet is routed to the corresponding
    output port, the latter is immediately marked as busy until the last character of the packet has
    passed through the switch indicated by its end\_of\_packet tail.
  \item If the requested output port is busy then the input port ceases to send flow control tokens
    to the source node, and thus halts the incoming packet until the output port becomes free.
    During this time, the link connecting the source node to the routing switch is blocked.
\end{itemize}

Wormhole routing is characterized by two properties: message contiguity (bits of different messages
cannot interleave) and minimal buffering (only few flits are buffered in intermediate nodes).
Contiguity and minimal buffering properties make for simple
hardware implementations, used in embedded systems such as satellites. For example, the bookkeeping at each
node is simplified because bits of different messages cannot be
interleaved. In addition, intermediate nodes can be made simple,
small, and simpler because the queues at each intermediate node
are only required to buffer few flits~\cite{RSSW94}.

\section{Network calculus}

Network calculus is based on min-plus algebra and convex
analysis~\cite{BCOQ,CDQV83,Roc70}. Min-plus operators such as
min-plus convolution and deconvolution are used to express and
handle constraints on data arrivals and service. Two important
notions in network calculus theory are \emph{arrival curves} and \emph{service curves}.
One of the main objectives of this theory is to calculate bounds on end-to-end delays and
data backlogs on servers. We give in this section a review on
basic results of network calculus.

Let $A(t), t\in\BN$ be a data arrival flow to a given server, such
that $A(t)$ is the cumulative arrival data up to time $t$. The
map $A$ is, by definition, non decreasing. We set, in addition,
$A(0)$ to $0$. In \emph{network calculus} theory, arrival and
service curves express constraints on
arrivals and services, and are used to determine performance bounds. A
minimal (resp. maximal) arrival curve for a flow $A(t), t\in\BN$
is any curve $\gamma(t)$ (resp. $\Gamma(t)$), $t\in\BN$ satisfying
$A(t)-A(s)\geq \gamma(t-s)$ (resp. $A(t)-A(s)\leq \Gamma(t-s)$),
$\forall 0\leq s\leq t$.

We can easily see that a curve $\Gamma$ is a maximal arrival curve
for $A$ if and only if $A\leq A*\Gamma$, where
$A*\Gamma(t)\stackrel{def}{=}\inf_{0\leq s\leq
t}[A(s)+\Gamma(t-s)]$. The operator $*$ is called min-plus
\emph{convolution} or simply convolution operator. An interesting
question is how one can chose a good maximal arrival curve among
from a set of arrival curves. A maximal arrival curve is defined
to bound arrivals of a given flow. Thus, a good maximum arrival
curve must give tight bounds at every time. Therefore we can at
least tell that if $\Gamma_1$ and $\Gamma_2$ are two maximum
arrival curves for a flow $A$ such that $\Gamma_1(t)\leq
\Gamma_2(t), \forall t\in\BN$, then $\Gamma_1$ is better than
$\Gamma_2$. Nevertheless, given a maximum arrival curve $\Gamma$,
a standard way  to find  a better  maximum arrival
curve is to compute its sub-additive closure.
 Let $\Gamma^{(n)}$ denote the curve defined by:
$\Gamma^{(0)}= I_0$, and
$\Gamma^{(n)}=\Gamma*\Gamma^{(n-1)}$, where $I_0(t)=+\infty,
\forall t>0$ and $I_0(0)=0$. Then we denote by $\Gamma^*$ the
curve $\Gamma^*=\inf_{n\geq 0}\Gamma^{(n)}$. It is easy to check
that $\Gamma^*(t)\leq \Gamma(t), \forall t\in\BN$; $\Gamma^*$ is
sub-additive; and $\Gamma^{**}=\Gamma^*$. The curve $\Gamma^*$ is
called the sub-additive closure of $\Gamma$.
Any maximum arrival curve $\Gamma$ can be 
replaced by its sub-additive closure  $\Gamma^*$ in the sense that
$\Gamma$ is a maximum arrival curve if and only if
 $\Gamma^*$ is a maximum arrival curve. Let $\Gamma_1$
and $\Gamma_2$ be two maximum arrival curves for a flow $A$, then
$\Gamma_1*\Gamma_2$ is also a maximum arrival curve for $A$. 
Actually, the best  maximum arrival
curve ({\it i.e.} the smallest one)
for a flow $A$, is the curve
$A\oslash A$, defined by $(A\oslash A)(t)=\sup_{s\geq
0}[A(t+s)-A(s)]$. Chang~\cite{Chang94} observed that $A\oslash A$ is
sub-additive. The operator $\oslash$ is called (min-plus)
\emph{deconvolution} operator.
Similarly, a curve $\gamma$ is a minimum arrival curve for $A$ if
and only if $A\leq A\bar{*}\gamma$, where
$A\bar{*}\gamma(t)\stackrel{def}{=}\sup_{0\leq s\leq
t}[A(s)+\gamma(t-s)]$. The operator $\bar{*}$ is called max-plus
convolution.

Let $A$ be an arrival flow to a given network node. We denote the
output flow from the node by $\bar{A}$. We say that the node
offers a minimum service curve $\omega$ if $\omega(0)=0$, $\omega$
is wide-sense increasing, and $\bar{A}\geq A*\omega$. The notion
of minimum service curve has its roots in the work of Parekh and
Gallager~\cite{PG93}. Service curves  can model links, servers, 
propagation delays, schedulers, regulators and window based on
throttles~\cite{Agr98}.
To give some intuition to the definition of
a minimum service curve, let us consider  the dynamics:
$\bar{A}(t)=\min\{A(t), \bar{A}(t-1)+e(t)\}, \forall t\geq 0$,
where $e(t)$ is given for all $t\geq 0$. Then, if we denote by
$E(t)=\sum_{s=0}^t e(s)$ and $\omega=E\oslash E$, then, $\omega$
satisfies $\bar{A}\geq A * \omega$. A minimum  service curve $\omega$ is {\it strict} if
during any data backlog period of
duration $u$ of the flow, the output flow is at least equal to
$\omega(u)$. It is not difficult to show that a minimum strict
service curve is also a minimum service curve;
see~\cite{Cha00,BT04}. Maximum service curves, and maximum strict
service curves are defined similarly.

Basic results of network calculus give bounds in the backlog, the delay and
the output burstiness on a server as functions of a given maximum
arrival curve $\Gamma$ for the input flow $A$ and a given minimum
service curve $\omega$ for the server.
\begin{itemize}
  \item We denote by $B(t)=A(t)-\bar{A}(t)$ the backlog at time $t$. Then the maximum backlog
    $B_{\max}$ is bounded by $B_{\max}\leq \sup_{s\geq 0}[\Gamma(s)-\omega(s)]$.
  \item We denote by $d_r(t)=\inf\{d\geq 0 \mid \bar{A}(t+d)\geq A(t)\}$ the virtual delay at
    time $t$. Then the maximum virtual delay $d_{\max}$ satisfies:
    $d_{\max}\leq \sup_{t\geq 0}\{\inf\{d\geq 0 \mid \omega(t+d)\geq \Gamma(t)\}\}$.
  \item Output burstiness: $\Gamma\oslash \omega$ is a maximum arrival curve for the output flow $\bar{A}$.
    If, in addition, a maximum service curve $\Omega$ is given, then the output flow
    $\bar{A}$ is constrained by the arrival curve
    $(\Gamma*\Omega)\oslash \omega$; see~\cite{Cru98}.
\end{itemize}

Simple but practical arrival and service curves are $(\sigma,\rho)$ arrivals and $(R,T)$ services.
A $(\sigma,\rho)$ arrival flow is an arrival flow constrained by the maximum arrival curve $\Gamma(t)=\sigma+\rho t$.
An $(R,T)$ server is a server that guarantees a minimum service curve $\omega(t)=R(t-T)^+$.
It is easy to check that for a $(\sigma,\rho)$ arrival flow served in
an $(R,T)$ server with $R > \rho$, one can  guarantee a maximum delay
$d=T+\sigma/R$, a maximum backlog $b=\sigma+\rho T$, and a $(\sigma+\rho T,\rho)$ output burstiness.

Arrivals to a given service node can be controlled using a window
flow control. In practice, buffers with limited sizes are used to
store data before serving it. The limit size of the buffers
constrains the service, and thus modify it. In a window flow
control server with window size $z$, data, once arrived, is
allowed to be served at time $t$ if the amount of data being in
the buffer is less than $z$. It is known that if $\omega$ is a
service curve for a server without buffering size limit
constraints, then the constrained server by a buffer of size $z$
guarantees a server curve $\omega * (I_z* \omega)^*$, where
$I_z(t)=+\infty, \forall t>0$, and $I_z(0)=z$. For example, if
$\omega(t)=R(t-T)^+$, then~\cite{Cha00,BT04}, if $z>RT$, then
$\omega* (I_z*\omega)^*=\omega$. That is to say that the buffer is
enough large not to constrain the server. In the case, where
$\omega<RT$, it is not difficult to check that
$(I_z*\omega)^*(t)\geq (z/T)t, \forall t\geq 0$, and thus $\omega*
(I_z*\omega)^*(t)\geq (z/T)(t-T)^+$; see for example~\cite{BT04}.

\section{Packetization}

In this section we introduce two new concepts: packet
operators, and packet curves. We give a short review in
packetization and some new results in non preemptive service, those
we use in the next section. The objective of the new formulations we make here is to
give a network calculus approach in calculating residual services
(and by this, delay and backlog bounds) in the case of serving
packet data flows under non preemptive packet service disciplines,
where packets may have different lengths.

We are concerned here by data arrival flows that arrive in packets.
Thus two flows can be distinguished: the flow of the amount of data
(bits) itself, independent of how it is clustered  in packets, which
we call simply data flow, 
and the flow of the number of packets, which we call the packet
flow. The idea here, is to define operators and minimum and
maximum curves that allow us to switch from the data flow space to
the packet flow space, and vice versa. This is similar to
packetization, but we will go one step further here
by introducing the constraints on packet lengths under the form of packet
curves.

The procedure is the following:
giving the service of a data flow, we deduce, first, the
service of the  packet flow, we serve packet flows (as
serving flows of data with equi-sized packets) and deduce residual
services of the packet flows, and finally, we deduce the residual
services of the data flows.
Packetizers describe how data is set in packets by an increasing
sequence of packet lengths~\cite{Cha00,BT04}. We replace this
sequence by a minimum and/or a maximum curves that  give the
minimum and/or the maximum number of packets in a given amount of
data. This new approach is more powerful than packetization and is
more in line  with the network calculus approach based on constraint curves.

For a wide-sense increasing function $f: \BR\to\BR$, the wide-sense increasing
functions $f^{-1}_-$ and $f^{-1}_+$, called respectively left and
right pseudo-inverses of $f$, are defined by:
$f^{-1}_-(x)=\inf\{t\in\BR, f(t)\leq x\}$ and
$f^{-1}_+(x)=\inf\{t\in\BR, f(t)\geq x\}$. 
Thus  $\forall
t\in\BR, f^{-1}_-\circ f(t)\leq t\leq f^{-1}_+\circ f(t)$,
$\forall x\in\BR, f\circ f^{-1}_-(x)\leq x\leq f\circ
f^{-1}_+(x)$,
and $\forall t\in\BR, (f^{-1}_-)^{-1}_-(t)\leq
(f^{-1}_+)^{-1}_-(t)\leq f(t)\leq (f^{-1}_-)^{-1}_+(t)\leq
(f^{-1}_+)^{-1}_+(t)$.

\begin{prop}\label{propo1}
  If $\gamma$ and $\Gamma$ are respectively minimal and maximal arrival curves for $A$
  then $A^{-1}_-(y)-A^{-1}_+(x) \leq \gamma^{-1}_+(y-x)$ and
  $A^{-1}_+(y)-A^{-1}_-(x) \geq \Gamma^{-1}_-(y-x)$.
\end{prop}
\proof We prove the first item and the proof is similar for the
second one. Let $0\leq x\leq y$. Let $s$ and $t$ be defined by
$s=A^{-1}_+(x)$ and $t=A^{-1}_-(y)$. Thus we have  $A(s)\geq x$
and $A(t)\leq y$. Then we get $y-x\geq A(t)-A(s)\geq
\gamma(t-s)=\gamma(A^{-}_-(y)-A^{-1}_+(x))$. then by applying
$\gamma^{-1}_+$, which is non decreasing, we obtain
$\gamma^{-1}_+(y-x)\geq A^{-1}_-(y)-A^{-1}_+(x)$, which gives the
result. \endproof

Let $A$ be an arrival data flow, with a maximum arrival curve
$\Gamma$, served in a server with a minimum service curve
$\omega$. The output of $A$ from the server is denoted by
$\bar{A}$. The virtual delay to the right at time $t$,
$d_r(t)=\inf\{d\geq 0 \mid
\bar{A}(t+d)\geq A(t)\}=\bar{A}^{-1}_-\circ A(t)-t$
is bounded by $d_r(t)\leq \sup_{t\geq 0}\left(\omega^{-1}_-\circ \Gamma
(t)-t\right)$. Similarly, the virtual delay to the left at time $t$, $d_l(t)=\inf\{d\geq 0
\mid A(t-d)\geq \bar{A}(t)\}=t-A^{-1}_-\circ \bar{A}(t)$ is bounded by 
$d\leq \sup_{t\geq 0}\left(t-\Gamma^{-1}_-\circ \omega(t)\right)$.

\subsection{Packet operators}

Packet and data operators
 will allow us to work with both data flows and 
packet flows, and in particular to switch from the data flow
context to the packet flow context or vice versa. Let $A$ be an
arrival flow, that is $A(t)$ gives the cumulated arrival data up
to time $t$. We define $\CP$ as the operator applied on data as
follows: For an amount $x$ of arrival data, $\CP(x)$ gives the
number of \emph{entire} packets in $x$. Thus the data contained in
$\CP(x)$ packets can eventually be less than $x$ (that is
$\CP^{-1}_-\circ\CP(x)\leq x$). The cumulated number of
entire packets arrived up to time $t$, denoted by $P(t)$ is simply
$\CP\circ A(t)$. Then $P=\CP\circ A$ is the arrival flow of the number
of packets of $A$.

Let us define $\CA$ over the domain  $\BN$ by $\CA(n)=\CP^{-1}_-(n)$
which is the data contained in $n$ packets. Then $\CA(n)_{n\in\BN}$
is a sequence of cumulative packet lengths, and is wide-sense
increasing. If we denote this sequence by $M$. Then the operator
$\CA\circ \CP=\CP^{-1}_-\circ\CP$ is an
$M$-packetizer~\cite{AR96,Cruz98,Cha00}.

By the same way as we bounded the service delay, where we have been placed on the time axis, we can now
be placed on the data axis, and bound the maximum packet length $L^{\max}$ associated to a given
operator packet $\CP$, as follows: $L^{\max}=\sup_{x\geq 0}(\CP^{-1}_+\circ \CP(x)-x)=
\sup_{x\geq 0}(x-\CP^{-1}_-\circ \CP(x))$.

Let $\CP$ be the packet flow associated to an arrival flow $A$, to
a given server, and let $\bar{A}$ denote the output flow of $A$
from the server. If the flow $A$ is served with First Come
First Served (FCFS) service~\footnote{We use the term \emph{FCFS}
to mean that the first arrived unit of data of one flow is the first served, while
we use the term \emph{FIFO} to mean that the first arrived data packet among from
packets of two or several data flows is the first served. So FIFO is a non preemptive
service policy.},
then the packet operator associated
to $\bar{A}$ is simply the packet operator associated to $A$. That
is, if we denote by $\Bar{\CP}$ the packet operator associated
to the output $\bar{A}$, then we have $\bar{\CP}=\CP$. That means
that a FCFS server do not repacketize data. We call this
assumption packetization invariance (PI) assumption under FCFS.
This assumption is realistic only when one data flow is served and
when the data is served with FCFS service. We will see below that
when more than one flow are served, the server often repacketize
data of the aggregate flow, depending on the applied service policy.
In the following, we recall a well-known result on packetization,
and rewrite it with our notations.
\begin{theo}{Packetization, \cite{BT04}.}\label{betahmin}
  If $\omega$ is a minimum service curve for $A$, then $t\mapsto (\omega(t)-l^{\max})^+$ is a minimum
  service curve for $\CP^{-1}_-\circ\CP\circ A$.
\end{theo}
\proof We have $\bar{A}(t)\geq A* \omega (t)$. Let $0\leq s\leq t$
such that $\bar{A}(t)\geq A(s)+\omega (t-s)$. Then
$\CP^{-1}_-\circ\CP\circ
\bar{A}(t)-\bar{\CP}^{-1}_-\circ\bar{\CP}\circ A(s) =
\CP^{-1}_-\circ\CP\circ \bar{A}(t)-\CP^{-1}_-\circ\CP\circ A(s)
\geq (\bar{A}(t)-l^{\max})-A(s)\geq \omega (t-s)-l^{\max}$. On the
other side, we have $\bar{A}(t)-A(s)\geq \bar{A}(s)-A(s)\geq
\omega(0)=0$, and since $\CP^{-1}_-\circ\CP$ is non decreasing, we
get $\CP^{-1}_-\circ\CP\circ\bar{A}(t)-\CP^{-1}_-\circ\CP\circ
A(s)\geq 0$.
\endproof

\subsection{Packet curves}

For a given arrival data flow $A$, one usually does not know the sequence
$A(t), t\in\BN$ for every $t$, but has  some statistical
information about $A$, namely the average in time of $A$, and the
maximal variance of $A$. This provides maximal arrival curves used 
to  compute performance bounds. Similarly, for an arrival
data flow, we are not always able to know exactly the associated
packet operator. However, we can have some information about
the maximum length of packets, the average length, and the distribution of
small and big packets on the data. With these informations we
define  minimum and maximum \emph{packet
curves}, that  give minimum and maximum numbers of packets in a
given amount of data.

\begin{defi}
  A curve $\pi$ (resp. $\Pi$) is said to be a minimum (resp. maximum) packet curve for $\CP$ if
  $\CP(y)-\CP(x)\geq \pi(y-x), \; \forall 0\leq x\leq y$
  (resp. $\CP(y)-\CP(x)\leq \Pi(y-x), \; \forall 0\leq x\leq y$).
\end{defi}

For example, the maximum packet length $l^{\max}$ and the average
packet length $L$ can be  expressed  using  the
minimum packet curve
$\pi$ as follows: $\pi(l^{\max})=1$ and $L=\lim_{x\to+\infty}
x/\pi(x)$. However, one  can have additionnal information, for
instance,   telling that,  in an
amount $x$ of data that is bigger than the maximum packet length
($x>l^{\max}$), there are at least $n$ packets with $1/l^{\max}< n/x
< 1/L$. A realistic example of a minimum packet curve $\pi$ is
$\pi(x)=\lfloor \max_{i\in\BN} R_i(x-T_i) \rfloor, \quad \forall x\in\BR_+$,
where $R_i, i\in\BN$ is a non decreasing real sequence, with
$R_0=1/l_{\max}$, and $T_i, i\in\BN$ is a non decreasing real
sequence, with $T_0=0$, and $R_1(l_{\max}-T_1)=1$.

\begin{exam}\label{expi}
  Here is  a simple example that will be  used throughout the paper
  for illustration purposes. Consider  a data arrival flow
  that arrives in packets. The packets are of lengths equal either to
  1 or 2 data units. In addition, in  three successive  packets, there is at least one packet of length 1, and at least one
  packet of length 2. The minimum and the maximum lengths are thus
  given by $l^{\min}=1$ and $\L^{\max}=2$.
  In this case, The curves $\pi, \pi^\star, \Pi$ and $\Pi^{-1}_-$ are shown in Figure~\ref{pi}.  
  
  \begin{figure}[htbp]
  \begin{center}
    \includegraphics[width=10cm]{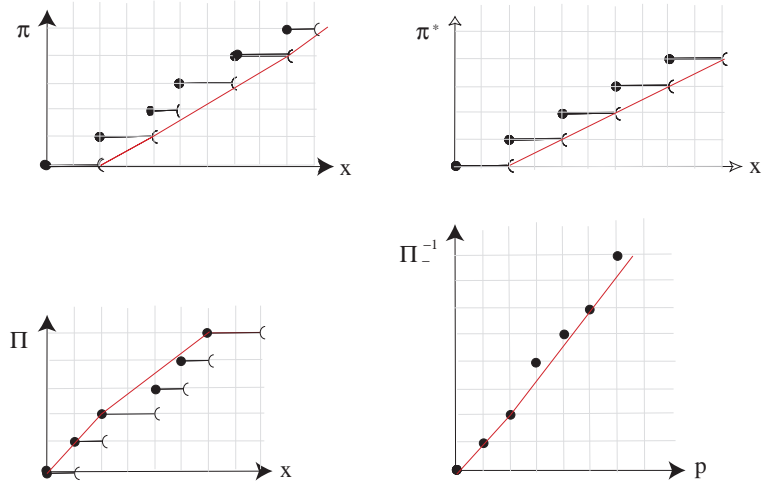}    
    \caption{The curves $\pi, \pi^*, \Pi$ and $\Pi^{-1}_-$. The continuous curves correspond to
      the piecewise affine curves that bound $\pi, \pi^*, \Pi$ and $\Pi^{-1}_-$; see (\ref{eqqq1})-(\ref{eqqq3}).}
    \label{pi}
  \end{center}
  \end{figure}

  The curves $\pi$ and $\Pi$ can be bounded respectively by piecewise
  affine curves. It is easy to check that:
  \begin{align}
    & \pi(x)\geq \max\left\{(1/2)\left(x-2\right)^+,(3/5)\left(x-(7/3)\right)^+\right\}, \label{eqqq1}\\
    & \Pi(p)\leq \min\left(x, (3/4)x+(1/2)\right), \label{eqqq2} \\
    & \Pi^{-1}_-(p)\geq \max\left\{p, (4/3)\left(p-(1/2)\right)^+\right\}. \label{eqqq3}
  \end{align}
\end{exam}

In the next sections, we will need the following results, whose proofs
are direct consequences of the definitions.

\begin{prop}\label{ba-1}
  Let $A$ be an arrival flow with packet operator $\CP$. We assume that $\CP$ is right-continuous.
  If $\Pi$ is a maximum packet curve for $\CP$, then $\forall 0\leq p\leq q\in\BN,
  \CP^{-1}_-(q)-\CP^{-1}_-(p)\geq \Pi^{-1}_-(q-p)$.
  If $\pi$ is a minimum packet curve for $\CP$, then $\forall 0\leq p\leq q\in\BN,
  \CP^{-1}_-(q)-\CP^{-1}_-(p)\leq \pi^{-1}_+(q-p)$.
\end{prop}
\proof
  Let $p,q\in\BN$ with $p\leq q$. Let $x=\CP^{-1}_-(p)$ and $y=\CP^{-1}_-(q)$. Since $\CP$ is right-continuous, we get
  $p=\CP(x)$ and $q=\CP(y)$.
  Then $q-p=\CP(y)-\CP(x)\leq \Pi(y-x)=\Pi(\CP^{-1}_-(q)-\CP^{-1}_-(p))$. Then by applying $\Pi^{-1}_-$ we get:
  $\Pi^{-1}_-(q-p)\leq \CP^{-1}_-(q)-\CP^{-1}_-(p)$, which gives the result. The proof is similar of the second item.
\endproof
\begin{prop}\label{alphah}~
   If $\Gamma$ is a maximum arrival curve for $A$, and $\Pi$ is a maximum packet curve for $\CP$,
   then $\Pi\circ \Gamma$ is maximum arrival curve for $P$.
   If $\gamma$ is a minimum arrival curve for $A$, and $\pi$ is a minimum packet curve for $\CP$,
   then $\pi\circ \gamma$ is minimum arrival curve for $P$.
\end{prop}
\proof $P(t)-P(s)=\CP\circ A(t)-\CP\circ A(s)\leq \Pi(A(t)-A(s))\leq \Pi\circ \Gamma (t-s)$. The proof is similar
  for the second item. \endproof

As for service curves, the curve $\CP\bar{\oslash}\CP$ is the best minimum packet curve for $\CP$, while
$\CP^{-1}_-\oslash \CP^{-1}_-$ is the best maximum packet curve for $\CA$.
Let $A$ be a data arrival flow with packet operator $\CP$, and let
$\pi$ be a minimum packet curve for $\CP$. The maximum length
$l_{\max}$ satisfies
$l_{\max}=\sup_{n\in\BN}(\CP^{-1}_-(n+1)-\CP^{-1}_-(n))=\CP^{-1}_-\oslash\CP^{-1}_-(1)\leq
\pi^{-1}_+(1)$.

\subsection{Minimum mean of service}

Served packetized data from time zero to time $t$ when $\bar{A}(t)=x$ is given by
$\CP^{-1}_-\circ\CP\circ \bar{A}(t)-\CP^{-1}_-\circ\CP\circ \bar{A}(0)=\CP^{-1}_-\circ\CP(x)$.
We will be interested here in the mean $\mu(X)$, in data, of the guaranteed
service after packetization, given, for a given level $X$ of data:
\begin{equation}
  \mu(X)=\frac{1}{X}\int_0^X \CP^{-1}_-\circ\CP(x)dx.
\end{equation}

Theorem~\ref{betahmin} bounds the guaranteed service after packetization:
$\forall t\geq 0, \CP^{-1}_-\circ\CP\circ\omega(t)\geq [w(t)-l^{\max}]^+$.
In the following, we give a result that bounds the mean $\mu(X)$.
\begin{theo}\label{theomean}
  If $\pi$ is a minimum packet curve for $\CP$, then the mean $\mu(X)$ of guaranteed service after packetization is bounded
  as follows: $\mu(X)\geq \frac{1}{X}\int_0^X \pi^{-1}_-\circ\pi(x)dx.$
\end{theo}
\proof
Let $L_i, i\in\BN$ be the lengths of the i-th arrival packets of $A$. That is:
$$\CP(x)=i,\quad \forall x \text{ satisfying } \sum_{j=1}^i L_j\leq x < \sum_{j=1}^{i+1} L_j.$$
Thus
$$\begin{array}{ll}
  S_1 & \stackrel{def}{=} \int_0^X (\CP^{-1}_+\circ\CP(x)-\CP^{-1}_-\circ\CP(x))dx, \\
      & = \sum_{j=1}^i \int_{\sum_{k=0}^{j-1}L_k}^{\sum_{k=0}^j L_k} (\CP^{-1}_+\circ\CP(x)-\CP^{-1}_-\circ\CP(x))dx
           + \int_{\sum_{k=0}^i L_k}^X (\CP^{-1}_+\circ\CP(x)-\CP^{-1}_-\circ\CP(x))dx, \\
      & = \sum_{j=1}^i L_j^2 + L_{i+1} (X-\sum_{j=1}^i L_j).
\end{array}$$
Similarly, we get:
$$S_2 \stackrel{def}{=} \int_0^X (\CP^{-1}_+\circ\CP(x)-x)dx = \frac{1}{2}\sum_{j=1}^i L_j^2 + L_{i+1}(X-\sum_{j=1}^i L_j)
          - \frac{1}{2}(X-\sum_{j=1}^i L_j)^2.$$
Hence
$$\frac{1}{2}X-\mu(X)=\frac{1}{X}\int_0^X(x-\CP^{-1}_-\circ\CP(x))dx
    =\frac{1}{X}(S_1-S_2)=\frac{1}{2X}\left(\sum_{j=1}^i L_j^2 + (X-\sum_{j=1}^i L_j)^2\right).$$
Now, let us order the packet lengths of $A$ in a non decreasing order, and use the notations:
$L'_1\geq L'_2\geq L'_3\geq ...$.

By definition of the curve $\pi$, starting from zero, and going ahead, the packet lengths are ordered
in non decreasing order, that is the order $L'_1\geq L'_2\geq L'_3\geq ...$. Thus, following the same steps
as above, by replacing $\CP$ with $\pi$, we get:
$$\frac{1}{2}X-\int_0^X \pi^{-1}_-\circ\pi(x)dx=\frac{1}{X}\int_0^X(x-\pi^{-1}_-\circ\pi(x))dx
    =\frac{1}{2X}\left(\sum_{j=1}^i {L'_j}^2 + (X-\sum_{j=1}^i L'_j)^2\right),$$
and since
$$\frac{1}{2X}\left(\sum_{j=1}^i {L'_j}^2 + (X-\sum_{j=1}^i L'_j)^2\right)\geq
   \frac{1}{2X}\left(\sum_{j=1}^i L_j^2 + (X-\sum_{j=1}^i L_j)^2\right),$$
we obtain
$$\mu(X)\geq \int_0^X \pi^{-1}_-\circ\pi(x)dx.$$
\endproof

Theorem~\ref{theomean} is used when we do not have the packet operator $\CP$ for $A$, but, instead, we have a
minimum packet curve $\pi$ for $\CP$. Note that if we do not have $\pi$ either, then we can only guarantee that
$\mu(X)\geq \frac{1}{X}\int_0^X(x-l^{\max})^+ dx=\left[\frac{1}{2}X-l^{\max}+\frac{1}{2} \frac{(l^{\max})^2}{X}\right]^+$.

\section{Non pre-emptive service}

We explain here how packet curves are used in non preemptive service. 
Suppose arrival flows are served under a given service discipline and with a given service curve.
The flows are assumed to arrive in packets of arbitrary lengths, and minimum and maximum packet curves are
supposed to be given. First, we determine a service curve for the aggregate flow of
packets arriving to the server, then we apply the service discipline
to the flow of packets. By this, we deduce residual services for the flows of packets. Finally we get the residual services for data flows.

Minplus convolution, power operation, and sub-additive closure operation are defined differently
for packet curves. Let $A_1$ and $A_2$ be two arrival flows with packet operators $\CP_1$ and $\CP_2$
respectively, and let $\pi_1$ and $\pi_2$ be minimal packet curves for $\CP_1$ and $\CP_2$ respectively.
We define $\CX_1$ and $\CX_2$ the sets $\CX_1=(\pi_1)^{-1}_-(\BN)$ and $\CX_2=(\pi_2)^{-1}_-(\BN)$,
and $\CX=\CX_1+\CX_2$. Operation $\star$ (minplus convolution for packet curves) is defined on packet curves as follows:
$$\forall x\in\BR_+, \pi_1\star\pi_2(x)=\left\{
  \begin{array}{ll}
    \min_{y\in\CX_1}[\pi_1(y)+\pi_2(x-y)], & \text{if } x\in\CX,\\
    (\pi_1\star\pi_2)(\max\{x'\in\CX, x'\leq x\}), & \text{otherwise}.
  \end{array} \right.$$
Similarly, $\pi^n$ is defined by
$\pi^0=I_0$ and $\pi^n=\pi\star\pi^{n-1}$ for $n\geq 1$, and $\pi^{\star}=\bigoplus_{n\geq 0}\pi^n$.
It is easy to check that any packet curve $\pi$ satisfies $\pi^\star\leq \pi$, and that $\pi^\star$ is sub-additive
on $\CX$, that is $\pi^\star(x+y)\leq \pi^\star(x)+\pi^\star(y), \forall x,y\in\CX$.
For example, if $\pi(x)=\lfloor x/2\rfloor$, then $\CX=2\BN$, and $\pi\star\pi=\pi=\pi^*$.
if $\pi(x)=\lfloor \max_{i}R_i(x-T_i)\rfloor$, with $R_0=1/l^{\max}$, then
$\pi^*(x)=\lfloor R_0 x\rfloor=\lfloor x/l^{\max}\rfloor$.

We suppose that the server offers a service curve $\omega$
(minimum or strict). We denote by $A$ the aggregate of arrival flows $A=A_1+A_2$,
and by $\CP$ the aggregate of packet flows $\CP=\CP_1+\CP_2$. The arrival flows of number of packets are denoted:
$P=\CP\circ A, P_1=\CP_1\circ A_1$ and $P_2=\CP_2\circ A_2$.
The outputs from the server are denoted by $\bar{A}, \bar{\CP}, \bar{P}, \bar{A}_1, \bar{A}_2, \bar{\CP}_1, \bar{\CP}_2, \bar{P}_1$ and $\bar{P}_2$ for respectively $A, \CP, P, A_1, A_2, \CP_1, \CP_2, P_1$ and $P_2$.

The packetization invariance (PI) assumption does not hold  here because the service of the two flows do not
necessarily preserve
the order of arrived packets. Indeed, the order of served packets depends on the service discipline.
To deal with this, we give  the following result.

\begin{prop}\label{routstrict}(Blind scheduling)
  If $\pi_1$ and $\pi_2$ are respectively packet curves for $\CP_1$ and $\CP_2$, then $(\pi_1\oplus\pi_2)^\star$ is
  a packet curve for $\bar{\CP}$.
\end{prop}
\proof We have to prove that
$\bar{\CP}(y)-\bar{\CP}(x)\geq (\pi_1\oplus\pi_2)^\star(y-x), \forall 0\leq x\leq y$.
Let $z_0, z_1, \ldots, z_n$, with $z_0=0$ and $z_n=x$, such that in any interval $(z_i,z_{i+1})$ of the cumulated output
data of $\bar{A}$, the output data corresponding to that interval is a data of only one flow among from the flows
$\bar{A}_1$ and $\bar{A}_2$. If we denote by $\CI_1, \CI_2\subset \{0,1,\ldots n\}$ such that
$i\in\CI_j, j=1,2$, when in $(z_i, z_{i+1})$, $\bar{A}$ is increased thanks to increasing of $\bar{A}_j$.

$$\bar{\CP}(z_{i+1})-\bar{\CP}(z_i)=
     \begin{cases}
       \bar{\CP}_1(z_{i+1})-\bar{\CP}_1(z_i) & \text{if } i\in\CI_1,\\
       \bar{\CP}_2(z_{i+1})-\bar{\CP}_2(z_i) & \text{if } i\in\CI_2.
     \end{cases}$$
Thus we have 
\begin{eqnarray*}
   \bar{\CP}(y)-\bar{\CP}(x) & = & \sum_{i\in\CI_1}\bar{\CP}_1(z_{i+1})-\bar{\CP}_1(z_i)
                                      + \sum_{i\in\CI_2}\bar{\CP}_2(z_{i+1})-\bar{\CP}_2(z_i),\\
                            & =    & \sum_{i\in\CI_1}\CP_1(z_{i+1})-\CP_1(z_i)
                                      + \sum_{i\in\CI_2}\CP_2(z_{i+1})-\CP_2(z_i),\\
                            & \geq & \sum_{i\in\CI_1}\pi_1(z_{i+1}-z_i) + \sum_{i\in\CI_2}\pi_2(z_{i+1}-z_i),\\
                            & \geq & \sum_{1\leq i\leq n} (\pi_1\oplus\pi_2)^\star(z_{i+1}-z_i)
                             \geq  (\pi_1\oplus\pi_2)^\star(y-x).
\end{eqnarray*}
\endproof

Note that if $\pi_1$ and $\pi_2$ are super-additive, then
$(\pi_1\oplus\pi_2)^\star(x)=\lfloor x/L^{\max}\rfloor\geq \frac{1}{L^{\max}}(x-L^{\max})^+$, where $L^{\max}$ is
the maximum packet length over all packets.

\begin{exam}
  Let $\pi_1(x)=\lfloor \max_{i\geq 0}R_{1i}(x-T_{1i})\rfloor$ and $\pi_2(x)=\lfloor \max_{i\geq 0}R_{2i}(x-T_{2i})\rfloor$.
  In this case, $\pi_1\oplus\pi_2$ takes also the form
  $(\pi_1\oplus\pi_2)(x)=\lfloor \max_{j\geq 0}R_{j}(x-T_{j})\rfloor$, with $R_0=\min(R_{10}, R_{20})$ and
  $T_0=\max(T_{10},T_{20})$. Thus, $(\pi_1\oplus\pi_2)^*(x)=\lfloor \min(R_{10},R_{20}) x\rfloor
   = \lfloor x/\max(l^{\max}_1,l^{\max}_2)\rfloor$.
\end{exam}

\subsection*{Service projectors}

We introduce  a new terminology that will ease the statement of  the main theorem of this section
(Theorem~\ref{betah}).
Let $P_1, P_2, \ldots, P_n$ be $n$ arrival flows to a server with a service curve $\omega$. We suppose that data
is measured and served in non decomposable units. A service discipline involves residual services for
the flows $P_i, 1\leq i\leq n$ with associated service curves $\omega_i, 1\leq i\leq n$ respectively.
Let us first note that strict service curves for packet flows are defined with respect to backlog periods
of the corresponding data flows.

\begin{defi}
  A curve $\omega$ is a strict service curve for a packet flow $P$ associated to a given data flow $A$, if in any
  backlog period $(s,t)$, \emph{with respect to $A$}, we have $\bar{P}(t)-\bar{P}(s)\geq \omega(t-s)$.
\end{defi}

\begin{defi}
  The maps $R_i, 1\leq i\leq n$ associating to the aggregate service curve $\omega$, the residual service curves
  $\omega_i, 1\leq i\leq n$ are called service projectors on flows $P_i, 1\leq i\leq n$, associated to the service
  discipline. $R_i: \omega\mapsto \omega_i$, and we have $\omega_i(t)=R_i\circ \omega (t), \forall t\geq 0$.
\end{defi}
We note here that in some projections, even though $\omega$ is a strict service curve for $A$,
$\omega_i=R_i\circ \omega$ may be minimum (not strict) service curves for $A_i, 1\leq i\leq n$.

\begin{theo}\label{betah}
  If $\omega$ is a strict service curve for $A$, and
  if $\pi_i$ and $\Pi_i, 1\leq i\leq n$ are minimum and maximum packet curves for $\CP_i, 1\leq i\leq n$ resp.,
  then $\left(\bigoplus_{i=1}^n \pi_i\right)^\star \circ \omega$ is a strict service curve for $P$,
  $R_i\circ\left(\bigoplus_{i=1}^n \pi_i\right)^\star \circ \omega, 1\leq i\leq n$ are service curves 
  (minimum or strict, depending on the projection) for $P_i, 1\leq i\leq n$ resp., and
  $(\Pi_i)^{-1}_-\circ R_i\circ\left(\bigoplus_{i=1}^n \pi_i\right)^\star \circ \omega$ are service curves
  (minimum or strict, depending on the projection) for $A_i$, respectively.
\end{theo}
\proof Let us prove the case where the projection does not preserve the service strictness.
The other case is easier.
\begin{itemize}
  \item From Proposition~\ref{routstrict}, we know that $(\oplus_{i=1}^n\pi_i)^\star$ is a packet curve for
    $\bar{\CP}$. That is $\bar{\CP}(y)-\bar{\CP}(x)\geq (\oplus_{i=1}^n\pi_i)^\star (y-x),
    \forall 0\leq x\leq y$.
    Let $(s,t)$ be a backlog period of $A$. We have $\bar{A}(t)-\bar{A}(s)\geq \omega(t-s)$.
    Then $\bar{P}(t)-\bar{P}(s)=\bar{\CP}\circ \bar{A}(t)-\bar{\CP}\circ \bar{A}(s)
    \geq (\oplus_{i=1}^n\pi_i)^\star(\bar{A}(t)-\bar{A}(s))\geq (\oplus_{i=1}^n\pi_i)^\star \circ \omega (t-s)$.
  \item By definition of $R_i, 1\leq i\leq n$, the curve $(\oplus_{i=1}^n\pi_i)^\star\circ \omega$ being a strict service
    curve for $P$ implies that $R_i\circ (\oplus_{i=1}^n\pi_i)^\star\circ \omega$ is a minimum service curve for $P_i$.
  \item According to Proposition~\ref{ba-1}, $\forall 1\leq i\leq n,
    (\CP_i)^{-1}_-(q)-(\CP_i)^{-1}_-(p)\geq (\Pi_i)^{-1}_-(q-p)$. Now, let $t\geq 0$.
    Since $R_i\circ (\oplus_{i=1}^n\pi_i)^\star\circ \omega$ is a minimum service curve for $P_i$, then 
    $\exists 0\leq s\leq t,
    \bar{P}_i(t)-P_i(s)\geq R_i\circ (\oplus_{i=1}^n\pi_i)^\star\circ \omega(t-s)$. Thus,
    $\bar{A}_i(t)-A_i(s)=(\bar{\CP}_i)^{-1}_-\circ \bar{P}_i(t)-(\CP_i)^{-1}_-\circ P_i(s)=
    (\CP_i)^{-1}_-\circ \bar{P}_i(t)-(\CP_i)^{-1}_-\circ P_i(s)\geq (\Pi_i)^{-1}_-(\bar{P}_i(t)-P_i(s))
    \geq (\Pi_i)^{-1}_-\circ R_i\circ\left(\bigoplus_{i=1}^n \pi_i\right)^\star \circ \omega(t-s)$. \endproof
\end{itemize}

\subsection{FIFO routing policy}
\label{fifo}

In wormhole routing, packets with different destination output ports
may arrive in the same input port of a given switch.
These packets are routed to their destination output ports following their arriving order. So FIFO
(First In First Out) policy
is applied on the level of input ports. Let us first recall a
basic result on FIFO routing service.

Let $A_1$ and $A_2$ be two arrival flows to a given server.
We denote by $\bar{A}_1$ and $\bar{A}_2$ respectively the outputs of $A_1$ and $A_2$ from that server.
Let $\Gamma_2$ be a maximal arrival curve for $A_2$, and $\omega$ a minimum service curve for
the aggregate flow. Let us denote by $\omega_{\theta}^1$ the family of functions $\omega_{\theta}^1=[\omega(t)-\Gamma_2(t-\theta)]^+\un_{\{t>0\}}$.

\begin{theo}{FIFO minimum service curve \cite{Cru98}}\label{review}
  If $A_1$ and $A_2$ are served under FIFO service policy, then we have
  $\bar{A}_1\geq A_1*\omega_{\theta}^1$ for all $\theta$,  and if $\omega_{\theta}^1$ is wide-sense increasing, then
  $\omega_{\theta}^1$ is a minimum service curve for $A_1$.
\end{theo}

\begin{coro}{$(R,T)$-minimum service curve for FIFO.}\label{corofifo}
  If $A_1$ and $A_2$ are served under FIFO service policy, with $\Gamma_i(t)=\sigma_i+\rho_i t, i=1,2$, and if
  $\omega(t)=R(t-T)^+$ is a minimum service curves for the server, then the curve
  $\omega_1(t)=(R-\rho_2)\left[t-\left(T+\sigma_2/R\right)\right]^+$
  is a minimum service curve for $A_1$, and thus the curve 
  $\bar{\Gamma}_1(t)=\left(\sigma_1+\rho_1 T+\sigma_2 \rho_1/R\right)+\rho_1 t$
  is a maximum arrival curve for~$\bar{A}_1$.
\end{coro}
\proof We can easily check that among from the service curves $\omega_{\theta}^1$ for $\theta\geq 0$, the
  $(R,T)$-minimum service curve that guarantees maximum of service, with respect to $\theta$, is attained
  for $\theta=T+\sigma_2/R$. \endproof

Now we consider the general case, where packets of one arrival flow may have different lengths. We suppose that
minimum packet curves $\pi_i, 1\leq i\leq n$ and maximum packet curves $\Pi_i, 1\leq i\leq n$ are associated to the flows
$A_i, 1\leq i\leq n$ respectively.

\begin{theo}\label{theofifo}
  If $\omega$ is a strict service curve for the aggregate flow, then a minimum service curve $\omega_i$ for
  flow $A_i$ is:
  $$\omega_i(t)=\Pi^{-1}_-\circ \left[\lfloor\omega(t)/L^{\max}\rfloor
    -\sum_{j\neq i}\Pi_j\circ \Gamma_j(t-\theta)\right]^+\un_{\{t\geq \theta\}}.$$
\end{theo}
\proof We just apply Theorem~\ref{betah}. Note that $(\bigoplus_{1\leq j\leq n}\pi_j)^*(x)=\lfloor x/L^{\max}\rfloor$,
  whenever $\pi_j, 1\leq j\leq n$ are super-additive. \endproof

\begin{exam}\label{ex1fifo}
  We consider here two flows $A_1$ and $A_2$ with maximum arrival
  curves $\Gamma_1(t)=\sigma_1+\rho_1t$ and
  $\Gamma_2(t)=\sigma_2+\rho_2t$ respectively.
  We assume that $\Pi(x)=\min(a_1 x, a_2 x+b)$ is a maximum packet curve for
  both packet operators $\CP_1$ and $\CP_2$ associated to $A_1$ and
  $A_2$ respectively. Thus we get $\Pi^{-1}_-(p)=\max((1/a_1)p,(1/a_2)(p-b)^+)$.
  We take $\omega(t)=R(t-T)^+$. Theorem~\ref{theofifo} gives
  $\omega_1$as follows:
  $$\omega_1(t)=\Pi^{-1}_-\circ \left[\lfloor\omega(t)/L^{\max}\rfloor
    -\Pi_2\circ \Gamma_2(t-\theta)\right]^+\un_{\{t\geq \theta\}}.$$
  Then we use the following simplifications:
  \begin{itemize}
    \item $\lfloor \omega(t)/L^{\max}\rfloor\geq \frac{1}{L^{\max}}(\omega(t)-L^{\max})^+
        = \frac{R}{L^{\max}}\left(t-(T+\frac{L^{\max}}{R})\right)^+$.
    \item $\Pi\circ\Gamma_2(t)=\min(a_1\sigma_2+a_1\rho_2 t, a_2\sigma_2+b+a_2\rho_2 t)$.
    \item In order to stay working with piecewise functions,
      $\theta$ is chosen as mentioned in Corollary~\ref{corofifo}.
      Then $\left[\lfloor\omega(t)/L^{\max}\rfloor -\Pi_2\circ \Gamma_2(t-\theta)\right]^+\un_{\{t\geq
      \theta\}}$ is bounded by
      $$\max\left\{
        \begin{array}{l}
          \left(\frac{R}{L^{\max}}-a_1\rho_2\right)\left[t-\left(T+\frac{L^{\max}}{R}+\frac{a_1\sigma_2L^{\max}}
            {R}\right)\right]^+,\\
          \left(\frac{R}{L^{\max}}-a_2\rho_2\right)\left[t-\left(T+\frac{L^{\max}}{R}+\frac{L^{\max}(a_2\sigma_2+b)}
            {R}\right)\right]^+
        \end{array}\right\}.$$      
    \item Then, applying $\Pi^{-1}_-$, we get:
  \end{itemize}    
    \begin{equation}
      \omega_1(t)=\max\left\{
      \begin{array}{l}
        \frac{1}{a_1}\left(\frac{R}{L^{\max}}-a_1\rho_2\right)
            \left[t-\left(T+\frac{L^{\max}}{R}+\frac{a_1\sigma_2L^{\max}}{R}\right)\right]^+,\\
        \frac{1}{a_1}\left(\frac{R}{L^{\max}}-a_2\rho_2\right)
            \left[t-\left(T+\frac{L^{\max}}{R}+\frac{L^{\max}(a_2\sigma_2+b)}{R}\right)\right]^+,\\
        \frac{1}{a_2}\left(\frac{R}{L^{\max}}-a_1\rho_2\right)
            \left[t-\left(T+\frac{L^{\max}}{R}+\frac{a_1\sigma_2L^{\max}}{R}+\frac{bL^{\max}}{R-a_1\rho_2L^{\max}}\right)\right]^+,\\
        \frac{1}{a_2}\left(\frac{R}{L^{\max}}-a_2\rho_2\right)
            \left[t-\left(T+\frac{L^{\max}}{R}+\frac{L^{\max}(a_2\sigma_2+b)}{R}+\frac{bL^{\max}}{R-a_2\rho_2L^{\max}}\right)\right]^+
      \end{array} \right\}.
    \end{equation}

\end{exam}

\subsection*{Link divergence}
\label{linkdiv}

\begin{figure}[htbp]      
  \begin{center}
    \includegraphics[width=6cm]{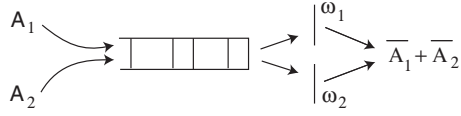}
    \caption{Link divergence.}
    \label{diverg1}
  \end{center}  
\end{figure}

Let us consider two arrival flows $A_1$ and $A_2$ to a given server. The flows arrive in packets through
a unique link, and the packets of both flows are served as one aggregate
arrival flow, but the service guaranteed for packets of each flow differs. We denote by $\omega_1$ and by $\omega_2$
service curves for packets of $A_1$ and $A_2$. Figure~\ref{diverg1} gives an illustration.
The question here is to determine a service curve for the aggregate flow $A_1+A_2$.

As in Proposition~\ref{routstrict}, minplus convolution, power operation, and sub-additive closure operation are redefined again
for service curves of packetized data. Let $A_1$ and $A_2$ be two arrival flows with packet operators $\CP_1$ and $\CP_2$
respectively, and let $\omega_1$ and $\omega_2$ be minimal service curves for $A_1$ and $A_2$ respectively.
We define $\CT_1$ and $\CT_2$ the sets $\CT_1=(\omega_1)^{-1}_-\circ(\pi_1)^{-1}_-(\BN)$ and $\CT_2=(\omega_2)^{-1}_-\circ(\pi_2)^{-1}_-(\BN)$,
and $\CT=\CT_1+\CT_2$. Operation $\diamond$ is then defined:
$$\forall t\in\BN, \omega_1\diamond\omega_2(t)=\left\{
  \begin{array}{ll}
    \min_{s\in\CT_1}[\omega_1(s)+\omega_2(t-s)], & \text{if } t\in\CT,\\
    (\omega_1\diamond\omega_2)(\max\{t'\in\CT, t'\leq t\}), & \text{otherwise}.
  \end{array} \right.$$
Similarly, $\omega^n$ is defined by
$\omega^0=I_0$ and $\omega^n=\omega\diamond\omega^{n-1}$ for $n\geq 1$, and $\omega^{\diamond}
   =\bigoplus_{n\geq 0}\omega^n$.
Similarly $\omega^\diamond\leq \omega$, and $\omega^\diamond$ is sub-additive
on $\CT$.

In the other side, we denote by $l^{\min}$ the minimum length of all packets, and we
define $T^{\max}=(\omega_1\oplus\omega_2)^{-1}_-(l^{\min})$. We assume in addition that $\omega_1$ and
$\omega_2$ are right-continuous, in such a way that we get $(\omega_1\oplus\omega_1)(T^{\max})=l^{\min}$.
Thus we have the following result.

\begin{prop}\label{propdiv}
  If $\omega_1$ and $\omega_2$ are strict service curves for $A_1$ and $A_2$ respectively
  then a strict service curve $\omega$ for $A_1+A_2$ is 
  $(\omega_1\oplus\omega_2)^\diamond$. Moreover, if $\omega_1$ and $\omega_2$ are
  super-additive, then $(\omega_1\oplus\omega_2)^\diamond=\lfloor(l^{\min}/T^{\max})t\rfloor
  \geq (l^{\min}/T^{\max})(t-T^{\max})^+$.
\end{prop}
\proof We can easily adapt the proof of Proposition~\ref{routstrict}
  to get $\bar{A}_1(t)+\bar{A}_2(t)-\bar{A}_1(s)-\bar{A}_2(s)
  \geq (\omega_1\oplus\omega_2)^\diamond (t-s)$.
  On the other side, if $\omega_1$ and $\omega_2$ are super-additive, then
  $(\omega_1\oplus\omega_2)^\diamond (t-s)\geq n (\omega_1\oplus\omega_2)^\diamond (T^{\max})
   = n (\omega_1\oplus\omega_2)(T^{\max})\geq \frac{t-s-\tau}{T^{\max}}\;(\omega_1\oplus\omega_2)(T^{\max})
  \geq \frac{l^{\min}}{T^{\max}}(t-s-T^{\max})$.
  Then since $\bar{A}_1(t)+\bar{A}_2(t)-\bar{A}_1(s)-\bar{A}_2(s)\geq
  0$, we obtain the result. \endproof
\begin{exam}\label{divrt}
  If $\omega_1(t)=R_1(t-T_1)^+$ and $\omega_2(t)=R_2(t-T_2)^+$,
  then $T^{\max}=\max(T_1+l^{\min}/R_1, T_2+l^{\min}/R_2)$.
\end{exam}

\subsection{Round-Robin service discipline}

Round-robin is a service policy that assigns service to each flow
in a circular order, \emph{without priority}. The order is
respected whenever possible; that is, if one flow is out of
packets, the next flow, following the defined order, takes its place. A
separate flow is considered for every data stream, and the server
serves a packet from any non-empty queue encountered, following a
cyclic order. When packets have variable sizes, flows with small
packets may be penalized.

Let $A_1, A_2,\cdots A_n$ be the $n$  arrival flows to the server.
Let $\Gamma_i, 1\leq i\leq n$ be maximum arrival curves
for $A_i, 1\leq i\leq n$, respectively. We assume that the flows arrive in packets with the
same size~$u$.

\begin{prop}
  If $\omega$ is a strict service curve for the aggregate flow $A$, then a minimum service curve $\omega_i$ for each
  flow $A_i$ is $\omega_i(t)=\max\{\frac{1}{n} (\omega(t)-nu), \omega(t)-\sum_{j\neq i}\Gamma_j(t),0\}$.
\end{prop}
\proof ~~
\begin{itemize}
  \item Firstly, it is known~\cite{BT04} that if $A_i, 1\leq i\leq n$ are served under ordered priority
    discipline,
    then the flow $i$ with the lowest priority guarantees a strict service curve
    $\left(\omega-\sum_{j\neq i}\Gamma_j\right)^+$. Secondly, it is trivial that under round-robin service, any flow
    $A_i$ guarantees a service better than the service it would guarantee if the flows are served under ordered priority,
    and if the flow has the lowest priority. Thus any flow $A_i$ guarantees a strict service at least equal to
    $\left(\omega-\sum_{j\neq i}\Gamma_j\right)^+$.
  \item With round-robin discipline, if $\omega$ is a strict service curve for the
    aggregate flow then $\frac{1}{n} [\omega(t)-nu]^+$ is a strict service curve for $A_i$. Indeed, in a given backlog
    period $(s,t)$, the worst case for flow $A_i$ is when $s$ corresponds to a time when
    a flow $A_i$ packet had just been served, and when $t$ corresponds to a time when a flow $A_i$ packet will be
    served just after $t$. That is, we lose one round. Also, in the worst case, the flows $A_j$ for $j\neq i$ are all
    backlogged in $(s,t)$. That is that the lost round is backlogged. Thus we lose $nu$ data.
    One round being lost, the flow $A_i$ guarantees $1/n$ times the remainder service. \endproof
\end{itemize}

Now we consider the general case, where packets of one arrival flow may have different lengths. we suppose that
minimum curves $\pi_i, 1\leq i\leq n$ and maximum curves $\Pi_i, 1\leq i\leq n$ are associated to the flows
$A_i, 1\leq i\leq n$ respectively.

\begin{theo}\label{theorr}
  If $\omega$ is a strict service curve for the aggregate flow, then a minimum service curve $\omega_i$ for
  flow $A_i$ is:
  $$\omega_i(t)=\max\left\{\frac{1}{n}(\Pi_i)^{-1}_- \circ ( \lfloor \omega(t)/L^{\max}\rfloor -1 ) ,
                 (\Pi_i)^{-1}_- \circ ( \lfloor \omega(t)/L^{\max}\rfloor -\sum_{j\neq i}\Pi_i\circ \Gamma_i),
                 0\right\}.$$
\end{theo}

\proof By applyin Theorem~\ref{betah} we get:
$$\omega_i(t)=(\Pi_i)^{-1}_- \circ \max\left\{\frac{1}{n}[(\bigoplus_{i=1}^n \pi_i)^* \circ \omega] -1,
                 (\bigoplus_{i=1}^n \pi_i)^*\circ \omega-\sum_{j\neq i}\Pi_i\circ \Gamma_i, 0\right\}.$$
Then since $(\Pi_i)^{-1}_-, 1\leq i\leq n$ are non decreasing, and
$(\bigoplus_{i=1}^n \pi_i)^* \circ \omega=\lfloor \omega(t)/L^{\max}\rfloor$ as long as $\pi_i, 1\leq i\leq n$
are super-additive, we get the result. \endproof

\begin{exam}\label{ex-rr}
  We consider here two flows $A_1$ and $A_2$ with maximum arrival
  curves $\Gamma_1(t)=\sigma_1+\rho_1t$ and
  $\Gamma_2(t)=\sigma_2+\rho_2t$ respectively.
  We assume that $\Pi(x)=\min(a_1 x, a_2 x+b)$ is a maximum packet curve for
  both packet operators $\CP_1$ and $\CP_2$ associated to $A_1$ and
  $A_2$ respectively. Thus we get $\Pi^{-1}_-(p)=\max((1/a_1)p,(1/a_2)(p-b)^+)$.
  We take $\omega(t)=R(t-T)^+$. Theorem~\ref{theorr} gives
  $\omega_1$as follows:
  $$\omega_1(t)=\max\left\{\frac{1}{2}(\Pi^{-1}_- \circ ( \lfloor \omega(t)/L^{\max}\rfloor -1 ) ,
                 \Pi^{-1}_- \circ ( \lfloor \omega(t)/L^{\max}\rfloor -\Pi\circ \Gamma_2),
                 0\right\}.$$
  The  following  piecewise affine bounds come from direct calculations:  
  \begin{itemize}
    \item $\lfloor \omega(t)/L^{\max}\rfloor\geq \frac{1}{L^{\max}}(\omega(t)-L^{\max})^+
        = \frac{R}{L^{\max}}\left(t-(T+\frac{L^{\max}}{R})\right)^+$.
    \item $\lfloor \omega(t)/L^{\max}\rfloor-1\geq \frac{R}{L^{\max}}\left(t-(T+\frac{2L^{\max}}{R})\right)^+$.
    \item $\frac{1}{2}\Pi^{-1}_-[\lfloor
      \omega(t)/L^{\max}\rfloor-1] \geq \max\left\{\frac{R}{2a_1L^{\max}}\left(t-\left[T+\frac{2L^{\max}}{R}\right]\right),
                                                   \frac{R}{2a_2L^{\max}}\left(t-\left[T+\frac{(2+b)L^{\max}}{R}\right]\right)\right\}$.
    \item $\Pi\circ\Gamma_2(t)=\min(a_1\sigma_2+a_1\rho_2 t, a_2\sigma_2+b+a_2\rho_2 t)$.
    \item $\lfloor \omega(t)/L^{\max}\rfloor-\Pi\circ\Gamma_2(t)\geq \max\left\{
       \begin{array}{l}
         \left(\frac{R}{L^{\max}}-a_1\rho_2\right)\left(t-\frac{a_1\sigma_2+a_1\rho_2(T+L^{\max}/R)}{R/L^{\max}-a_1\rho_2}\right)^+,\\
         \left(\frac{R}{L^{\max}}-a_2\rho_2\right)\left(t-\frac{a_2\sigma_2+b+a_2\rho_2(T+L^{\max}/R)}{R/L^{\max}-a_2\rho_2}\right)^+
       \end{array}\right\}$.
  \end{itemize}
  Then we obtain
  \begin{equation}\label{eqrr}
    \omega_1(t)=\max\left\{
       \begin{array}{l}
         \frac{1}{a_1}\left(\frac{R}{L^{\max}}-a_1\rho_2\right)
                        \left(t-\frac{a_1\sigma_2+a_1\rho_2(T+L^{\max}/R)}{R/L^{\max}-a_1\rho_2}\right)^+,\\
         \frac{1}{a_1}\left(\frac{R}{L^{\max}}-a_2\rho_2\right)
                        \left(t-\frac{a_2\sigma_2+b+a_2\rho_2(T+L^{\max}/R)}{R/L^{\max}-a_2\rho_2}\right)^+,\\
         \frac{1}{a_2}\left(\frac{R}{L^{\max}}-a_1\rho_2\right)
                        \left(t-\frac{a_1\sigma_2+a_1\rho_2(T+L^{\max}/R)}{R/L^{\max}-a_1\rho_2}
                        - b/\left(\frac{R}{L^{\max}}-a_1\rho_2\right)\right)^+,\\
         \frac{1}{a_2}\left(\frac{R}{L^{\max}}-a_2\rho_2\right)
                        \left(t-\frac{a_2\sigma_2+b+a_2\rho_2(T+L^{\max}/R)}{R/L^{\max}-a_2\rho_2}
                        - b/\left(\frac{R}{L^{\max}}-a_2\rho_2\right)\right)^+
       \end{array}\right\}.
  \end{equation}
\end{exam}

\section{A wormhole binary switch model}

In the following, we present a wormhole switch model, and determine residual services and delays of several
flows passing through the switch. This work is done in the framework of a spacewire network study, where
wormhole routing is applied. Therefore, our model will be based on spacewire switch characteristics.
A spacewire switch comprises a number of spacewire link interface (encoder-decoders) and a routing matrix.
The routing matrix enables the transfer of packets arriving at one link to another link interface on the
routing switch. In practice, each link interface can be considered as comprising an input port (link
interface receiver) and an output port (link interface transmitter). In the model we present here, we
separate receivers from transmitters in order to explain the modeling. In practice, either only path
addressing, or a combination of several addressings are implemented. We suppose here that switches
implement only path addressing. Flows are distinguished by their destination addresses, which are, in the
case of path addressing, a sequence of output ports.

When a packet arrives  in a routing switch, the corresponding output port is determined. The output port
does not transmit any other packet until the packet that is currently transmitted is sent. In our model,
a routing switch is given by a number of input ports, a number of output ports, and a routing matrix.
The routing matrix associates to each input port all possible output port destinations.

\subsubsection*{FIFO routing for input ports}

Packets with different destination output ports arrive by one link to a given input port. There is no service delay
on the input ports. The latter only provide the routing of packets to associated output ports. However, each packet
must wait until its destination output port is available. Moreover,
packets arriving to one input port are routed under the
FIFO routing policy to their associated output ports. Therefore, if two packets
with  different destinations arrive in sequence in an input
port, and if the first  packet waits for its output port to be available,
then the second  packet must also wait.

\begin{figure}[htbp]  
  \begin{center}
    \includegraphics[width=7cm]{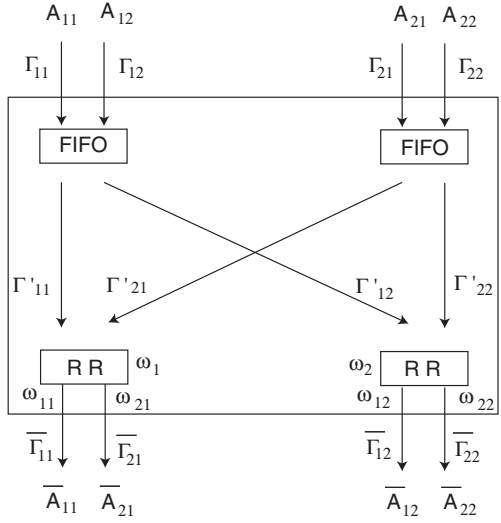}
    \caption{Wormhole switch modeling.}
    \label{switch}
  \end{center}
\end{figure}

\subsubsection*{Round Robin service in output ports}

Routed packets to a given output port are served under the round-robin discipline. The discipline is
used in all output ports. That is to say that connected input ports to a given output port are
served following a given cyclic order.

\subsection{Piecewise linear calculus}

Although no service delay is considered in input ports, the arrival
flows  $A_{ij}$ are modified before arriving
to the output ports. This is due to the FIFO routing imposed on the input ports. Indeed, when $A_{12}$ is being
served on the output port $O_2$, $A_{11}$ cannot be served by the outport $O_1$ even if the latter is free.
So arrivals of $A_{11}$ to the output port $O_1$ are not the same as arrivals of $A_{11}$ to the input port
$I_1$. Here is how this is taken into account:
Let $\Gamma_{ij}, i,j\in\{1,2\}$ be maximum arrival curves for $A_{ij}, i,j\in\{1,2\}$, respectively, and let
$\omega_1$ and $\omega_2$ be strict service curves for the servers of the output ports $O_1$ and $O_2$; see
Figure~\ref{switch}.

We assume that the arrivals $\Gamma'_{ij}, i,j\in\{1,2\}$ to the output ports
are given and are $(\sigma,\rho)$ arrivals. We use the notations: $\Gamma'_{ij}(t)=\sigma'_{ij}+\rho_{ij} t$
(the rates $\rho_{ij}$ stay unchanged). Then the residual services $\omega_{ij}, i,j\in\{1,2\}$ are computed
according to the round-robin discipline. After this, we deduce the output burstinesses
$\bar{\Gamma}_{ij}, i,j\in\{1,2\}$, which are $(\sigma,\rho)$ arrival curves. We use the notations
$\bar{\Gamma}_{ij}=\bar{\sigma}_{ij}+\rho_{ij} t$. Note that the variables here are $\sigma'_{ij}, i,j\in\{1,2\}$.
We obtain:
$\bar{\sigma}_{ij}=f_{ij}(\sigma'_{11},\sigma'_{12},\sigma'_{21},\sigma'_{22}), i,j\in\{1,2\}$.

On the other side, the FIFO routing at the input ports is taken into
account as follows. 
$A_{11}$ and $A_{12}$ arrive to the input port $I_1$, and are served respectively with services $\omega_{11}$
at the output port $O_1$, and $\omega_{12}$ at the output port
$O_2$. We apply the result from 
\emph{link divergence}, and obtain a strict service curve
$\omega'_1$ for the aggregate flow 
$A_{11}+A_{12}$. Then we include the buffering limit constraint on the input port $I_1$, and deduce a new minimum
service curve of the aggregate flow $A_{11}+A_{12}$. Then we apply the result on FIFO routing to determine minimum
service curves for flows $A_{11}$ and $A_{12}$. These service curves
are denoted by $\bar{\omega}_{11}$ and
$\bar{\omega}_{12}$. This gives the output burstinesses
$\bar{\Gamma}_{11}$ and $\bar{\Gamma}_{12}$.
Similarly, the output burstinesses   $\bar{\Gamma}_{21}$ and
$\bar{\Gamma}_{22}$ can be computed.  Thus
$\bar{\sigma}_{ij}, i,j\in\{1,2\}$ are given  as functions of the variables $\sigma'_{ij}, i,j\{1,2\}$:
$\bar{\sigma}_{ij}=g_{ij}(\sigma'_{11},\sigma'_{12},\sigma'_{21},\sigma'_{22}), i,j\in\{1,2\}$.
Finally we solve, in $\sigma'$, the system
\begin{equation}
  f_{ij}(\sigma'_{11},\sigma'_{12},\sigma'_{21},\sigma'_{22})
  = g_{ij}(\sigma'_{11},\sigma'_{12},\sigma'_{21},\sigma'_{22}), \quad i,j\in\{1,2\}.
\end{equation}
We give the details below.

\subsubsection*{Round-robin effect}

Applying round-robin discipline, we obtain the residual services (strict service
curves) $\omega'_{ij}, i,j\in\{1,2\}$ as follows (we apply
formula~(\ref{eqrr}))~:
\begin{equation}\label{omegapij}
  \omega'_{ij}(t)=\max\left\{R'_{ij1}(t-T'_{ij1})^+, R'_{ij2}(t-T'_{ij2})^+,R'_{ij3}(t-T'_{ij3})^+,
    R'_{ij4}(t-T'_{ij4})^+\right\}.
\end{equation}
For example, $\omega'_{11}$ is given by:
\begin{align*}
  & R'_{111}=\frac{1}{a_1}\left(\frac{R_1}{L^{\max}}-a_1\rho_{21}\right), \quad
    R'_{112}=\frac{1}{a_1}\left(\frac{R_1}{L^{\max}}-a_2\rho_{21}\right),\\
  & R'_{113}=\frac{1}{a_2}\left(\frac{R_1}{L^{\max}}-a_1\rho_{21}\right), \quad
    R'_{114}=\frac{1}{a_2}\left(\frac{R_1}{L^{\max}}-a_2\rho_{21}\right),\\
  & T'_{111}=\frac{a_1\sigma'_{21}+a_1\rho_{21}(T_1+L^{\max}/R_1)}{R_1/L^{\max}-a_1\rho_{21}}, \quad
    T'_{112}=\frac{a_2\sigma'_{21}+b+a_2\rho_{21}(T_1+L^{\max}/R_1)}{R_1/L^{\max}-a_2\rho_{21}},\\
  & T'_{113}=\frac{a_1\sigma'_{21}+a_1\rho_{21}(T_1+L^{\max}/R_1)}{R_1/L^{\max}-a_1\rho_{21}}
         -b/\left(\frac{R_1}{L^{\max}}-a_1\rho_{21}\right), \\
  & T'_{114}=\frac{a_2\sigma'_{21}+b+a_2\rho_{21}(T_1+L^{\max}/R_1)}{R_1/L^{\max}-a_2\rho_{21}}
         -b/\left(\frac{R_1}{L^{\max}}-a_2\rho_{21}\right).
\end{align*}

Then the output burstinesses $\bar{\Gamma}_{ij}, i,j\in\{1,2\}$ are given by:
\begin{equation}\label{burst1}
  \bar{\sigma}_{ij}=f_{ij}(\sigma')=\sigma'_{ij}+\rho_{ij} \min\{T'_{ij1},T'_{ij2},T'_{ij3},T'_{ij4}\}, \quad i,j\in\{1,2\}.
\end{equation}

\subsubsection*{Link divergence effect}

We determine a strict service curve $\omega'_1$ for the
aggregate flow $A_{11}+A_{12}$, and a strict service curve $\omega'_2$ for the
aggregate flow $A_{21}+A_{22}$. For this, we use the curves $\omega'_{11}$ and
$\omega'_{12}$ given in formula~(\ref{omegapij}), and apply
Proposition~\ref{propdiv} and Example~\ref{divrt}. The curves
$\omega'_1$ and $\omega'_2$ are thus given as follows:
\begin{align}
  & \omega'_1(t)=\frac{l^{\min}}{T^{\max}_1}(t-T^{\max}_1)^+, \label{omegap1}\\
  & \omega'_2(t)=\frac{l^{\min}}{T^{\max}_2}(t-T^{\max}_2)^+, \label{omegap2}
\end{align}
where
$T^{\max}_1=(\omega'_{11}\oplus\omega'_{12})^{-1}_-(l^{\min})$ and
$T^{\max}_2=(\omega'_{21}\oplus\omega'_{22})^{-1}_-(l^{\min})$.
According to formula~(\ref{omegapij}), we obtain:
\begin{align*}
  & T^{\max}_1=\max\left\{\begin{array}{l}
           \min\{l^{\min}/R'_{111}+T'_{111},l^{\min}/R'_{112}+T'_{112},l^{\min}/R'_{113}+T'_{113},
                  l^{\min}/R'_{114}+T'_{114},\},\\
           \min\{l^{\min}/R'_{121}+T'_{121},l^{\min}/R'_{122}+T'_{122},l^{\min}/R'_{123}+T'_{123},
                  l^{\min}/R'_{124}+T'_{124},\}
         \end{array}\right\}. \\
  & T^{\max}_2=\max\left\{\begin{array}{l}
           \min\{l^{\min}/R'_{211}+T'_{211},l^{\min}/R'_{212}+T'_{212},l^{\min}/R'_{213}+T'_{213},
                  l^{\min}/R'_{214}+T'_{214},\},\\
           \min\{l^{\min}/R'_{221}+T'_{221},l^{\min}/R'_{222}+T'_{222},l^{\min}/R'_{223}+T'_{223},
                  l^{\min}/R'_{224}+T'_{224},\}
           \end{array}\right\}.
\end{align*}

\subsubsection*{Buffering limit effect}

\begin{figure}[htbp]
  \begin{center}
    \includegraphics[width=8cm]{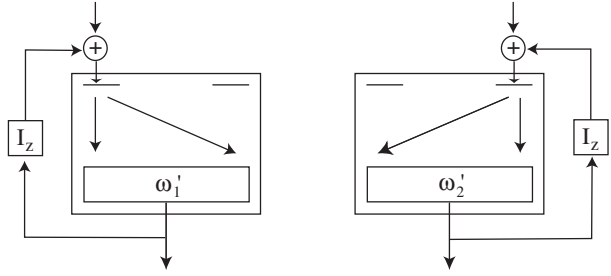}    
    \caption{Buffering.}
    \label{buf}
  \end{center}
\end{figure}

A limited size buffer is considered on each input port. To take this constraint into account, we use
window flow control results presented above. An illustration is given on Figure~\ref{buf}.
Here, the curves $\omega'_1$ and $\omega'_2$ are used, since, only the aggregate flow on each input port
counts. We denote by $\bar{\omega}_1$ and $\bar{\omega}_2$ the service curves of the aggregate
flows $A_{11}+A_{12}$ and $A_{21}+A_{22}$ respectively, that takes into account the buffering limit size.
The curves are obtained $\bar{\omega}_1=\omega'_1*(I_z*\omega'_1)^*$ and
$\bar{\omega}_2=\omega'_2*(I_z*\omega'_2)^*$, where $z$ is the size of the buffers.

The curves $\omega'_1$ and $\omega'_2$ are given as $(R,T)$ service curves, in equations~(\ref{omegap1})
and~(\ref{omegap2}). We consider here the non trivial case where $z<l^{\min}$.
In this case, the curves $\bar{\omega}_1$ and $\bar{\omega}_2$ are
obtained as follows:
\begin{align}
  & \bar{\omega}_1(t)=\frac{z}{T_1^{\max}}\;(t-T_1^{\max})^+, \\
  & \bar{\omega}_2(t)=\frac{z}{T_2^{\max}}\;(t-T_2^{\max})^+.
\end{align}

\subsubsection*{FIFO effect}

The arrival flows $A_{11}$ and $A_{12}$ arrive, now, to a given
server with a strict service curve $\bar{\omega}_1$, and are served
under FIFO discipline. Similarly, the arrival flows
$A_{21}$ and $A_{22}$ arrive to a server with a strict service curve
$\bar{\omega}_2$, and are served under FIFO discipline.
Then we can use Theorem~\ref{theofifo} and Example~\ref{ex1fifo}
given in section~\ref{fifo}. For example, the effective service
curve $\omega_{11}$ is given by:
$\omega_{11}(t)=\max\{R_{111}(t-T_{111})^+, R_{112}(t-T_{112})^+,R_{113}(t-T_{113})^+,R_{114}(t-T_{114})^+\}$,
where
\begin{align*}
  & R_{111}=\frac{1}{a_1}\left(\frac{z/T_1^{\max}}{L^{\max}}-a_1\rho_{12}\right), \quad
    R_{112}=\frac{1}{a_1}\left(\frac{z/T_1^{\max}}{L^{\max}}-a_2\rho_{12}\right),\\
  & R_{113}=\frac{1}{a_2}\left(\frac{z/T_1^{\max}}{L^{\max}}-a_1\rho_{12}\right), \quad
    R_{114}=\frac{1}{a_2}\left(\frac{z/T_1^{\max}}{L^{\max}}-a_2\rho_{12}\right),\\
  & T_{111}=T_1^{\max}+\frac{L^{\max}}{z/T_1^{\max}}+\frac{a_1\sigma_{12}L^{\max}}{z/T_1^{\max}}, \quad
    T_{112}=T_1^{\max}+\frac{L^{\max}}{z/T_1^{\max}}+\frac{L^{\max}(a_2\sigma_{12}+b)}{z/T_1^{\max}},\\
  & T_{113}=T_1^{\max}+\frac{L^{\max}}{z/T_1^{\max}}+\frac{a_1\sigma_{12}L^{\max}}{z/T_1^{\max}}
               +\frac{bL^{\max}}{z/T_1^{\max}-a_1\rho_{12}L^{\max}},\\
  & T_{114}=T_1^{\max}+\frac{L^{\max}}{z/T_1^{\max}}+\frac{L^{\max}(a_2\sigma_{12}+b)}{z/T_1^{\max}}
               +\frac{bL^{\max}}{z/T_1^{\max}-a_2\rho_{12}L^{\max}}.
\end{align*}
$\omega_{ij}, i,j\in\{1,2\}$ are given similarly:
\begin{equation}\label{omegafin}
  \omega_{ij}(t)=\max\{R_{ij1}(t-T_{ij1})^+, R_{ij2}(t-T_{ij2})^+,R_{ij3}(t-T_{ij3})^+,R_{ij4}(t-T_{ij4})^+\},
\end{equation}
where $R_{ijk}$ and $T_{ijk}, i,j\in\{1,2\}, k\in\{1,2,3,4\}$ are obtained similarly.

The output burstinesses $\bar{\Gamma}_{ij}, i,j\in\{1,2\}$ are then given again by:
\begin{equation}\label{burst2}
  \bar{\sigma}_{ij}=g_{ij}(\sigma')=\sigma_{ij}+\rho_{ij} \min\{T_{ij1},T_{ij2},T_{ij3},T_{ij4}\}, \quad i,j\in\{1,2\}.
\end{equation}

Finally, we have to solve, in $\sigma'$, the system \{(\ref{burst1}), (\ref{burst2})\}, that is:
\begin{equation}\label{sys1}
  f_{ij}(\sigma')=g_{ij}(\sigma'), \quad i,j\in\{1,2\}.
\end{equation}
This can be done by solving numerically the following fixed point problem:
\begin{equation}\label{sys2}
  \sigma'_{ij}=\sigma_{ij}+\rho_{ij} \min\{T_{ij1},T_{ij2},T_{ij3},T_{ij4}\}
                     -\rho_{ij} \min\{T'_{ij1},T'_{ij2},T'_{ij3},T'_{ij4}\}, \quad i,j\in\{1,2\}.
\end{equation}

Once the vector $\sigma'$ is obtained, the effective services of the arrivals $A_{ij}, i,j\in\{1,2\}$, through
the switch, are simply $\omega_{ij}, i,j\in\{1,2\}$, given in formula~(\ref{omegafin}). Maximum delays $d_{ij}$ for
flows $A_{ij}, i,j\in\{1,2\}$ are then obtained~:
\begin{equation}
  d_{ij}=\min_{k=1,2,3,4}\{T_{ijk}+\sigma_{ij}/R_{ijk}\}, \quad i,j\in\{1,2\}.
\end{equation}

\begin{exam}(Symmetric case)
  Let us consider the symmetric case, where the arrivals $A_{ij}, i,j\in\{1,2\}$ are constrained by the same
  arrival curve $\Gamma(t)=\sigma+\rho t$, and have the same maximum packet curve $\Pi(x)=\min(a_1x,a_2x+b)$;
  both output ports guarantees the same service $R(t-T)^+$, and the switches on the input ports have the same
  size $z$. For $\Pi$, we take the curve given in Example~\ref{expi},
  that is $a_1=1/10, a_2=3/40$ and $b=1/20$. We have also from the same example $l^{\min}=10$ and $L^{\max}=20$.
  
  $(\sigma,\rho)$--$(R,T)$ calculus with $\rho=1, T=2$ gives
  the effective service for the flows $A_{ij}, i,j\in\{1,2\}$, which is the same for
  all these flows, as follows:
  
\begin{center}
  \begin{tabular}{|c||c|c|c||c|c|c|}
    \hline
    Parameters & $\sigma$ & $R$ & $z$ & $\sigma'$ & $\bar{\sigma}$ & obtained maximum delay \\
    \hline
    ref. case & 3 & 7 & 8 & 106.69 & 136.93 & 129.16 \\
    \hline
    new $R$ & 3 & 8 & 8 & 51.43 & 64.19 & 54.99 \\
    \hline
    new $z$ & 3 & 7 & 9 & 63.14 & 81.50 & 72.91\\
    \hline
    new $\sigma$ & 2 & 7 & 8 & 73.43 & 94.60 & 89.16\\
    \hline
  \end{tabular}
\end{center}
\end{exam}

\subsection{Feedforward networks}

We show in this section how to extend this approach to feedforward networks composed of buffered nodes (source nodes,
switches, and destination nodes), and links. The connection of different nodes by the links define several
virtual paths. We are interested here in the calculation of the end-to-end delay through a given virtual path.
For this we need to compute the service through the virtual path. A virtual path is defined simply by a sequence
of a source node, zero, one or several switches, indicating the input and the output ports used, and a destination
node.

The procedure is to compute the service through each switch, without taking into account buffering limits,
compose all the services of the switches corresponding to the path, and finally include the buffering limit available
through the whole path, by adding the sizes of all the buffers through the path. The composition of services corresponding to
the switches of the path, is done algebraically. That is, if $\omega_1, \omega_2, \ldots, \omega_n$ are minimum service
curves of $n$ successive servers, then $\omega_1*\omega_2*\ldots *\omega_n$ is a minimum service curve through the
path~\cite{Cha00,BT04}.
Then, we simply apply the window flow control presented above to take into account the buffering limit. We show this
on a small feedforward network.

\begin{exam}
  On Figure~\ref{vpath} we take a feedforward network with two source nodes $A$ and $B$, four switches
  $S_1, S_2, S'_1$ and $S'_2$ and two destination nodes $D_1$ and $D_2$.
  We denote by $A_{ijk}$ (respectively $B_{ijk}$) the flow of data going from the source node $A$
  (respectively $B$) to the destination node $D_k$, through switches $S_i$ and $S'_j$. We are
  interested in computing the maximum delay for the flow $A_{221}$, that is for messages going from
  the source node $A$ to the destination node $D_1$ through the switches $S_2$ and $S'_2$.
  On Figure~\ref{bufpath}, we show the scheme of the calculation of the service through the buffered virtual path.

  \begin{figure}[htbp]
    \begin{center}
      \includegraphics[width=10cm]{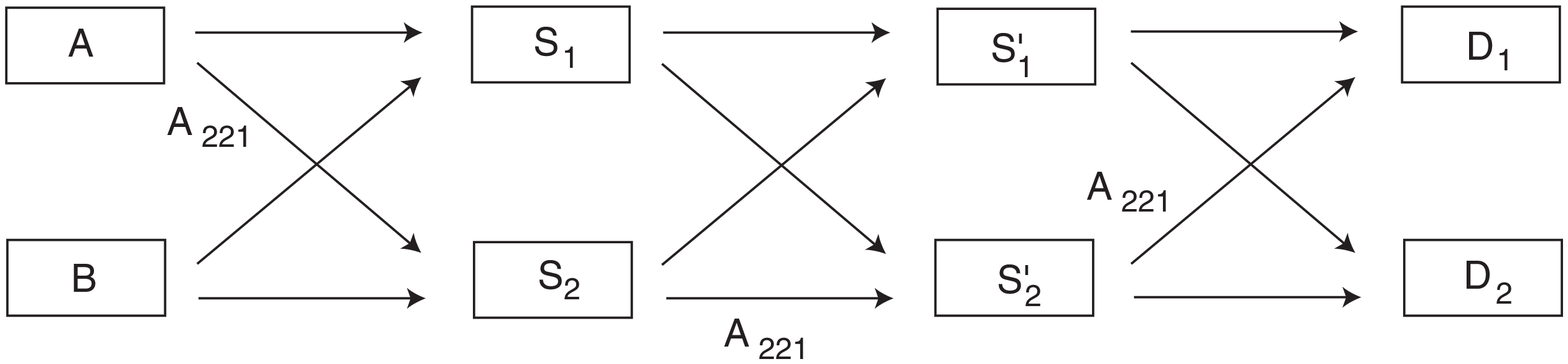}
      \caption{Network calculus.}
      \label{vpath}
    \end{center}
  \end{figure}

  Let us consider the symmetric case, where the arrivals $A_{ijk}$ and $B_{ijk}$ have the same maximum arrival curve
  $\Gamma(t)=\sigma+\rho t$, and where the output ports of the switches $S_1,S_2,S'_1$ and $S'_2$ guarantees
  the same strict service curve $\omega(t)=R(t-T)^+$. We follow the procedure explained above to compute
  the service of $A_{211}$ through a non buffered switch, then we compose the service through the switch $S_2$
  with the service through the switch $S'_2$ (which are the same because of the symmetry), and finally,
  we take into account the buffering, by using the window flow control technic with a window size equal to $2z$
  (two buffers of size $z$). In order to satisfy $z<Rt$, we have to chose $z<7.86$. We took here $z=6$.
  Finally we obtain the following results:
\begin{center}
  \begin{tabular}{|c||c|c|c||c|}
    \hline
    Parameters & $\sigma$ & $R$ & $z$ & obtained maximum delay \\
    \hline
    ref. case & 3 & 7 & 6 & 134.14 \\
    \hline
    new $R$ & 3 & 8 & 6 & 79.34 \\
    \hline
    new $z$ & 3 & 7 & 7 & 131.78\\
    \hline
    new $\sigma$ & 2 & 7 & 6 & 102.78\\
    \hline
  \end{tabular}
\end{center}  
  
\begin{figure}[htbp]
  \begin{center}
    \includegraphics[width=7cm]{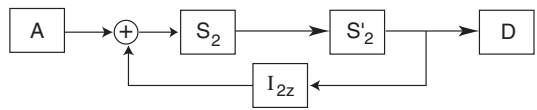}
    \caption{A buffered virtual path.}
    \label{bufpath}
  \end{center}
\end{figure}
\end{exam}


\bibliographystyle{plain}
\bibliography{biblio}

\end{document}